\documentclass[prb,twocolumn,superscriptaddress,floatfix,showpacs,preprintnumbers,amssymb,amsmath,a4]{revtex4}

\usepackage[latin1]{inputenc}
\usepackage{graphicx}% Include figure files
\usepackage{dcolumn}% Align table columns on decimal point
\usepackage{bm}% bold math
\usepackage[mathscr]{eucal}
\usepackage{epsfig}
\usepackage{rotating}

\begin{document}

\newcommand{\beq}{\begin{equation}}
\newcommand{\eeq}{\end{equation}}
\newcommand{\ket}{\rangle}
\newcommand{\bra}{\langle}
\newcommand{\A}{\mathbf{A}}
\preprint{ }
\title{Spectroscopy of three strongly coupled flux qubits}

\author{A. O. Niskanen}
\email{niskanen@frl.cl.nec.co.jp}
\affiliation{CREST-JST, Kawaguchi, Saitama 332-0012,Japan}
\affiliation{VTT Information Technology, Microsensing, POB 1207, 02044 VTT, Finland}

\author{K. Harrabi}
\affiliation{CREST-JST, Kawaguchi, Saitama 332-0012,Japan}

\author{F. Yoshihara}
\affiliation{The Institute of Physical and Chemical Research (RIKEN), Wako, Saitama 351-0198, Japan}

\author{Y. Nakamura}
\affiliation{CREST-JST, Kawaguchi, Saitama 332-0012,Japan}

\affiliation{The Institute of Physical and Chemical Research (RIKEN), Wako, Saitama 351-0198, Japan}
\affiliation{NEC Fundamental Research Laboratories, Tsukuba, Ibaraki 305-8501, Japan}

\author{J.S. Tsai}
\affiliation{CREST-JST, Kawaguchi, Saitama 332-0012,Japan}

\affiliation{The Institute of Physical and Chemical Research (RIKEN), Wako, Saitama 351-0198, Japan}
\affiliation{NEC Fundamental Research Laboratories, Tsukuba, Ibaraki 305-8501, Japan}

\date{\today}

\begin{abstract}
We have carried out spectroscopic measurements of a system of three strongly coupled four-junction flux qubits.
The samples studied cover a wide range of parameters with the coupling energy between neighboring qubits varying
between 0.75 GHz and 6.05 GHz. 
The observed complicated spectra agree well with eight-level theory. The experiments are relevant for the realization 
of a tunable coupling between qubits.
\end{abstract}


\maketitle

The potential realization of a full scale quantum computer requires the ability of
coupling multiple qubits together preferably so that the coupling can be turned on and
off at will. In the context of Josephson junction qubits there is a number of promising
theoretical suggestions \cite{bertet, flicforq, grajcar2} 
as well as already several experiments with coupled qubits \cite{pashkin,yamamoto, berkley,martinis,4qub,delftspec,ploeg,steffen}.
In order for a quantum computer to be truly scalable it must be 
possible to couple many qubits together without degrading the coherence time severely.
Tunable coupling has not been so far demonstrated in an experiment where the coherence time
would be equally good as in the case of so-called optimally biased qubits\cite{saclay,delft2,yale,us}.
A problem common with many coupling methods is that in order to realize a two-qubit gate the biases need to
be switched away from the region in the parameter space where the decoherence is minimal. In other words, 
while the coupling between two (or more) qubits is strengthened, the coupling between the qubits and the environment is,
in many cases, also strengthened.

This paper describes spectroscopic experiments on three strongly coupled flux qubits,
which is relevant for instance for the scheme suggested in Ref.~\onlinecite{mypaper}. In that scheme the parametric coupling of two 
detuned optimally biased flux qubits is realized through the microwave modulation of their tunable mutual inductance 
realized using a third qubit. 
Thus, the system we study is a set of three antiferromagnetically coupled flux qubits.
When the flux threading the qubits is near half flux quantum (modulo flux quantum $\Phi_0$) the system of three qubits is 
reasonably well described by the Hamiltonian
\beq\label{eq:ham}
H=-\frac{1}{2}\sum_{j=1}^3\left(\Delta_j\sigma_x^j+\epsilon_j\sigma_z^j \right)+\sum_{k=1}^3\sum_{l=k+1}^3J_{kl}\sigma_z^k\sigma_z^l,
\eeq
where $\epsilon_j=2I_{{\rm p}j}(\Phi_j-\Phi_0/2)$ is the energy bias of qubit $j$ controllable through the flux $\Phi_j$
threading the qubit loop.  Near the half-flux-quantum point each qubit experiences a double well potential
and the tunneling energy through the potential barrier separating the wells is $\Delta_j$. The wells correspond to currents of magnitude 
$I_{{\rm p}j}$ circulating in opposite directions along the loop and the above Hamiltonian is actually written in this circulating current basis.
The antiferromagnetic interaction between the qubits $k$ and $l$ is characterized by the coupling strength $J_{kl}=M_{kl}I_{{\rm p}k}I_{{\rm p}l}$
where $M_{kl}$ is the mutual inductance.

\begin{figure}
\begin{picture}(150,200)
\put(-30,0){\includegraphics[width=0.38\textwidth]{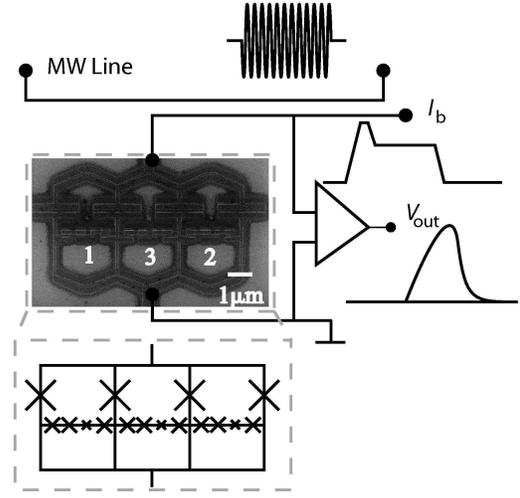}}
\end{picture}
\caption{\label{fg:sem}
SEM image of the sample and the principle of the measurement.
}
\end{figure}

A sample realizing such a system is shown in Fig.~\ref{fg:sem}. 
The interesting part of the sample consists of three four-Josephson-junction flux qubits,
similar to those studied individually in Refs.~\onlinecite{delft2}~and~\onlinecite{us}, 
coupled together by sharing one edge of the superconducting loop.  
The qubits 1 and 2 are thus expected to be weakly coupled due to small geometric inductance
whereas each of them is expected to be strongly coupled to the qubit 3 through kinetic inductance. Each qubit has one smaller
Josephson junction whose area is about $\alpha\approx0.5$ times smaller than that of the larger junctions having areas of  about 200 nm by 400 nm.
The qubits are coupled to a four-junction readout SQUID consisting of the three larger loops. The SQUID can be 
seen as a dc SQUID whose junctions are replaced by dc SQUIDs. This design was chosen in order to reduce the 
influence of the inevitable noise in the bias current.
The coupling of the qubits to the SQUID is also through kinetic inductance. 
Since the different states of the qubits correspond to different magnetic field configurations in the SQUID loops
we expect a finite population of the excited states to result in a small but detectable change of the switching current of the SQUID.

The samples used in this study were fabricated out of Al using standard shadow evaporation through a Ge enforced mask patterned 
with e-beam lithography. 
The designs of samples A, B and C 
were slightly different. In sample A we used an on-chip capacitor to shunt the SQUID\cite{delft,delft2,us}.
In samples B and C this capacitor was removed in order to clean up the measurable spectrum from resonances such as LC resonances and 
the plasma resonance.
An additional benefit of removing the large superconducting shunt capacitor was improved flux stability.
Both designs had no on-chip bias resistors but instead we used an on-chip LC filter consisting of a long superconducting line\cite{grenoble}
and a superconducting capacitor. The two plates of the Al parallel plate capacitor were defined in a separate e-beam lithography step
and the insulator was formed by heavily oxidizing the bottom layer.
The estimated cutoff of the fliter was 100--200 MHz. In sample A the inductor was about 15 mm long and 600 nm wide resulting in a transmission
line resonance around 10 GHz. The shorter line of samples B and C (3.2 mm) gives a resonance around 20 GHz.
The oxidation of the junction was done using a mixture of O$_2$ (10\%) and Ar (90\%) between the depositions of the 20 nm and 30 nm Al layers.
For the microwave line and bonding pads we used evaporated gold film defined by optical lithography. Part of the microwave 
line also had an Al layer. 
The measurements were carried out in a dilution refrigerator with a base temperature around 20 mK. The fridge had a triple $\mu$-metal 
magnetic shield. 

The principle of the measurement\cite{delft} is shown in Fig.~\ref{fg:sem}. First a microwave pulse of
typically 5 $\mu$s duration is applied to the microwave line on chip. Then immediately after this a current pulse of about 5--10 ns is applied
to the SQUID. The height is chosen so that the SQUID switches about every second time. The current pulse is followed by a trailing plateau
whose height is about {70\%} of the switching current. The application of current through the SQUID changes the magnetic fluxes 
experienced by the qubits and this shift is assumed to be adiabatic. The purpose of the plateau is to maintain the SQUID in the voltage state
if and only if the SQUID switched. By repeating each measurement typically $10^4$ times and 
counting the relatively slow voltage pulses (few $\mu$s) we can deduce the switching probability $P_{\rm sw}$ 
under particular circumstances.
To supply the flux bias to the qubits we used an external coil capable of inducing about 20 Gs.
To carry out spectroscopy on the qubits we first located them in the flux space. The basic measurement was then carried out by sweeping the 
microwave frequency and magnetic field and repeating the above mentioned measurement scheme. At each flux point the 
height of the current pulse was adjusted to get roughly the same $P_{\rm sw}$ in the absence of the microwave.
The qubits (as well as other resonances) cause deviations from this probability enabling the characterization of the spectrum 
of the three-qubit system.
\begin{figure}

\centerline{\includegraphics[width=0.5\textwidth]{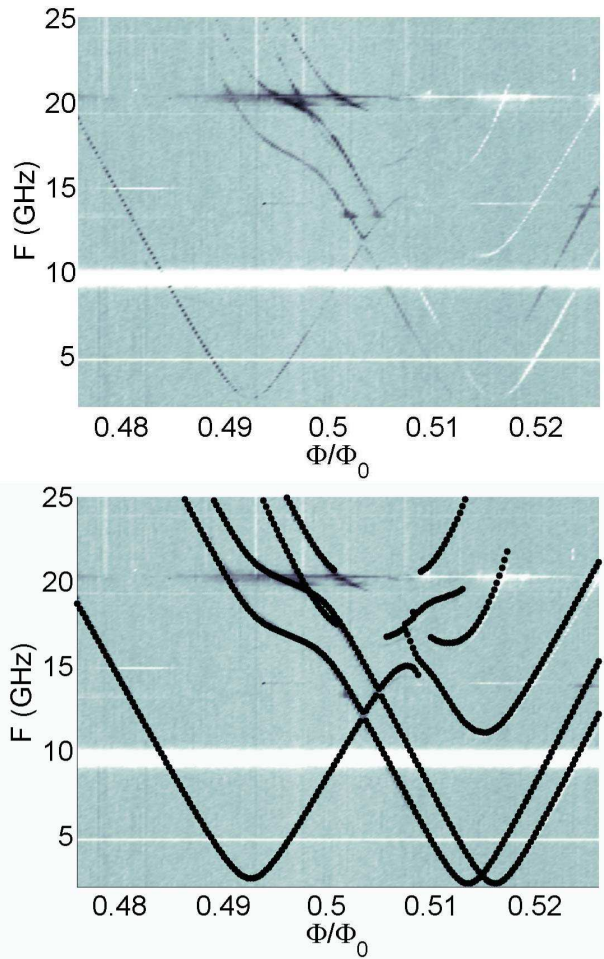}}

\caption{\label{fg:060328}
Spectroscopy of sample A. Top panel is an intensity plot of $P_{\rm sw}$
while the bottom panel includes the numerically calculated spectrum. Black dots denote excitation
energies from the ground state $|0\ket$ for the states $|\Psi\ket$ with $|\bra 0|\sum_{j=1}^3\sigma_z^j| \Psi\ket|^2\geq 0.002$.
In the calculation we used $\Delta_1/h=2.2$ GHz, $\Delta_2/h=2.2$ GHz, $\Delta_3/h=2.5$ GHz,
$J_{13}/h=J_{23}/h=2.05$ GHz, $J_{12}=0$ GHz,
$\Delta\Phi_1/\Phi_0 =-0.0128$, $\Delta\Phi_2/\Phi_0 =-0.0102$, $\tilde{I}_{\rm p1}=194$ nA, 
$\tilde{I}_{\rm p2}=194$ nA, $\tilde{I}_{\rm p3}=179$ nA.
The lines shared by the qubits have the dimensions of 1.35 $\mu$m$\times$30 nm$\times$160 nm.
Qubits 1 and 2  have $\alpha=0.5$ while qubit 3 has $\alpha=0.475$.
Oxidation was done with 25 mTorr for 5 min. 
Switching measurement yields $I_{\rm c,max}^{\rm SQUID}\approx 8.8\,\mu$A. }
\end{figure}

\begin{figure}
\centerline{\includegraphics[width=0.5\textwidth]{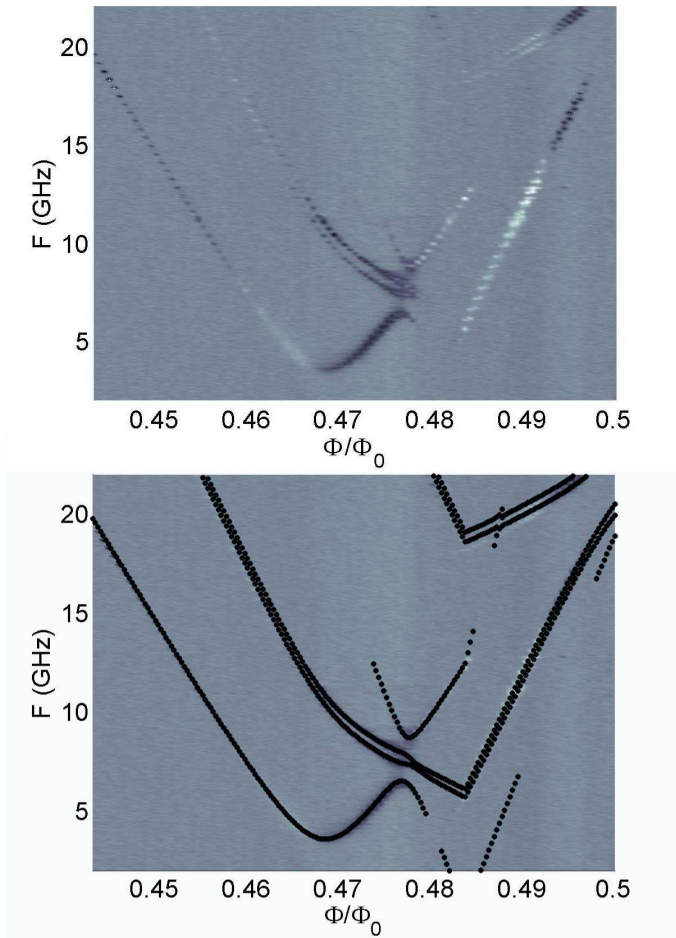}}
\caption{\label{fg:060801_3}
Spectroscopy of sample B. The parameters used in the calculation are 
$\Delta_1/h=1.3$ GHz, $\Delta_2/h=1.0$ GHz, $\Delta_3/h=3.6$ GHz, 
$J_{13}/h=J_{23}/h=6.05$ GHz, $J_{12}=0$ GHz,
$\Delta\Phi_1/\Phi_0 =-0.0096$, $\Delta\Phi_2/\Phi_0 =  -0.010$, $\tilde{I}_{\rm p1}=156$ nA, $\tilde{I}_{\rm p2}=156$ nA, $\tilde{I}_{\rm p3}=124$ nA.
We only show levels for which $|\bra 0|\sum_{j=1}^3\sigma_z^j| \Psi\ket|^2\geq 0.0001$.
The lines shared by the qubits have the dimensions of 1.35 $\mu$m$\times$20 nm$\times$100 nm.
Qubits 1 and 2 have $\alpha=0.5$ while qubit 3 has $\alpha=0.45$. 
Oxidation was done with 25 mTorr for 10 min. We get $I_{\rm c,max}^{\rm SQUID}\approx20.2\,\mu$A.
}
\end{figure}

\begin{figure}
\centerline{\includegraphics[width=0.5\textwidth]{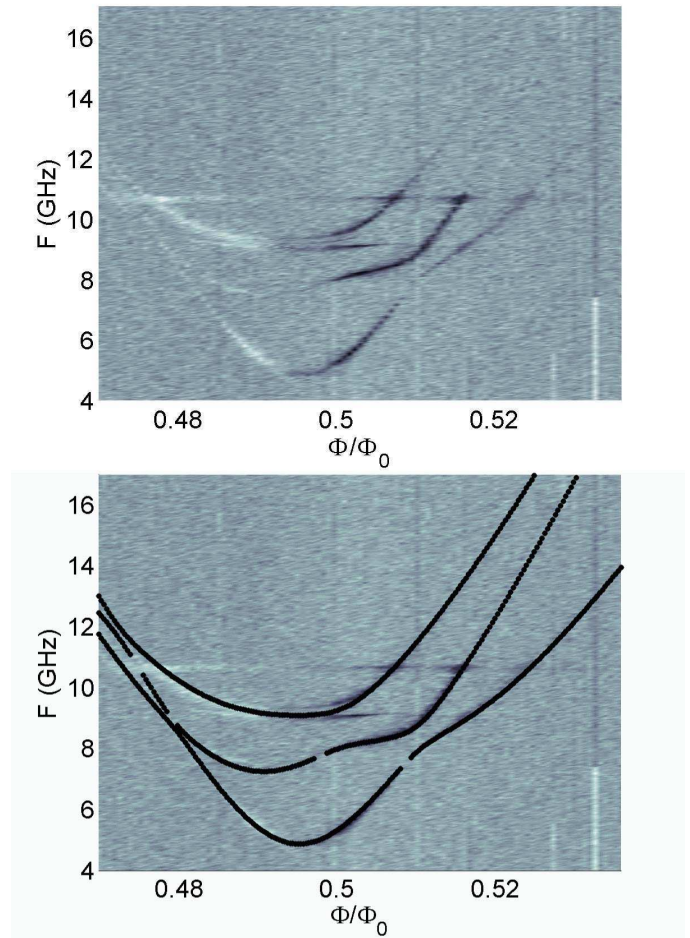}}
\caption{\label{fg:060818_3}
Spectroscopy of sample C. In the calculation we used
$\Delta_1/h=5.0$ GHz, $\Delta_2/h=7.5$ GHz, $\Delta_3/h=8.2$ GHz, 
$J_{13}/h=0.79$ GHz, $J_{13}/h=0.755$ GHz, $J_{12}=0$ GHz,
$\Delta\Phi_1/\Phi_0 =-0.0041$, $\Delta\Phi_2/\Phi_0 =  -0.0087$, $\tilde{I}_{\rm p1}=81$ nA, $\tilde{I}_{\rm p2}=78$ nA, $\tilde{I}_{\rm p3}=64$ nA.
Levels with $|\bra 0|\sum_{j=1}^3\sigma_z^j| \Psi\ket|^2\geq 0.005$ are shown.
The different slopes are due to the fact that
the qubits 1 and 2 were designed with  $\alpha=0.5$ and $\alpha=0.475$ while the qubit 3 has $\alpha=0.45$.
The lines shared by the qubits have the dimensions of 1.35 $\mu$m$\times$20 nm$\times$100 nm.
Oxidation was done with 35 mTorr for 10 min. We get $I_{\rm c,max}^{\rm SQUID}\approx5.4\,\mu$A.
}
\end{figure}

Figure~\ref{fg:060328} is an example of such a spectroscopic measurement on sample A. The horizontal lines are resonances due to the presence
of on-chip capacitors. The strong resonance around 10 GHz is most likely a half-wavelength resonance in the long bias line. 
However, the qubits are clearly visible and form a rich spectrum. 
The top panel shows the measured spectrum only while in the bottom panel a theoretical calculation of the spectrum is shown
on top of the measured spectrum.
The calculation is simply done numerically using Eq.~\ref{eq:ham} by finding the excited state eigenvalues
and subtracting the ground state eigenvalue from them. It is worthwhile stressing that the theoretical part is not based on 
a fitting procedure such that there may be considerable amount of error in all the parameters, but 
the agreement does seem at least qualitatively very good. The persistent currents denoted by $\tilde{I}_{{\rm p}k}$ in the figure captions
are the actual values used in the computation and do not account for the shielding effect. 
The actual persistent currents 
are expected to be significantly larger and also the mutual inductance is expected to be smaller than what it seems considering the 
currents $\tilde{I}_{{\rm p}k}$.  Based on the measured normal state resistivity  of $\rho^{Al}_{4.2\rm K}=11\,\mu\Omega$cm 
we estimate using\cite{terhaar}  $L_{\rm kin}= (\Phi_0eR_{\rm n})/(\pi^2\Delta_{\rm BCS})$ that
$M_{13}=M_{23}=31$ pH. In order for this to be consistent with the coupling energy the shielding should be about 11\%.
Due to the coupling we observe more than three levels
and everywhere the spectrum lines cannot be simply associated with a particular qubit. Roughly speaking, the line
having a minimum of about 2.5 GHz corresponds to the qubit 3 and the other two minima around 2.2 GHz correspond to the qubits 1 and 2.
This can be justified by noting that the qubit 3 is the only one coupled strongly to two other qubits.
The slightly higher tunneling energy is because the qubit 3 was fabricated with $\alpha=0.475$ as the designed 
ratio of the large and small junction area while qubits 1 and 2 had $\alpha=0.5$.
Since we are using only one control magnet we can write the flux biases as $\Phi_j=\Phi+\Delta\Phi_j$. The small offsets
$\Delta\Phi_j$ are, in the absence of trapped vortices, due to slight differences in the areas. Since we only work in rather 
small range around $\Phi_0/2$ the effect of area difference on the slope is not significant. We choose a convention where $\Delta\Phi_3=0$.
Note however that due to the interaction between qubits, not even the qubit 3 has minimum exactly at $\Phi_0/2$. The lack of visibility of
some levels at certain field can be understood as small transition matrix element. In fact, in the theoretical calculation we have plotted
only levels $|\Psi\ket$ for which $|\bra 0|\sum_{j=1}^3\sigma_z^j| \Psi\ket|^2\geq 0.002$. 

Figures~\ref{fg:060801_3}~and~\ref{fg:060818_3} illustrate measured spectra for samples B and C. As can be seen the 
calculations again agree extremely well with the measurement. Together with sample A the samples cover a very wide range of parameters.
It is especially noteworthy that the coupling energy can be made as large as 6 GHz without using coupling junctions \cite{4qub}.
The shielding effect as well as the coupling inductance are expected to be very large in these samples since the line shared by the qubits and SQUID
is made thinner and narrower than in sample A.
A prediction based on normal state resistance yields the estimate $M_{13}=M_{23}=74$ pH.
In order to explain the observed coupling energy the shielding in sample B should be about 40\%
and in sample C about 14\%. These numbers are reasonable although hard to verify. The scaling of these percentages is 
correct as a function of $I_{\rm c,max}^{\rm SQUID}$.

Even though the predicted kinetic inductances are large one may suspect that the large coupling could be 
partially due to shared Josephson inductance $L_{\rm J}$.
Namely, a fraction of the persistent current of each qubit could flow through the large junctions of the SQUID.
However, this effect may only reduce the coupling from the prediction based on shared kinetic inductance 
since the effect of the large shared  $L_{\rm J}$ is cancelled
by the fact that the fraction of the persistent currents flowing through the big junctions decreases with increasing  $L_{\rm J}$. 
Explicitily, if the currents of two qubits $k$ and $l$ can flow through two inductances $2L_1$ (kinetic)
and $2L_2$ (kinetic plus Josephson), and if half of these inductances are shared with the other qubit, then 
$J_{kl}=L_1L_2/(L_1+L_2)I_{{\rm p}k}I_{{\rm p}l}$. This clearly is maximal for $L_2\to \infty$.
An additional experimental point related to this is that in all samples 
we reproduced the measured spectrum well by setting the direct coupling of the qubits 1 and 2 to zero even though their currents
have a chance to flow through a ``shared'' large Josephson junction.

In conclusion we have carried out spectroscopic measurements of three strongly coupled flux qubits. We demonstrated
that one can achieve a wide range of coupling strengths using kinetic inductance for the coupling.
The relatively high coupling energy is attributable to the rather large normal state resistivity of our aluminum film.
For instance in  Ref.~\onlinecite{terhaar} the resistivity seems to be about 7 times smaller while e.g.~in 
Ref.~\onlinecite{jukka} it is about half of our value.
We find that the effective three-qubit Hamiltonian describes the measured spectrum well.
Based on our experience it is possible to reproducibly fabricate qubits whose flux biases differ by less than 1\%
even when using only a single bias coil.
The different effective area of qubit 3 in all samples is due to the fact that its position is not symmetric with respect to other qubits. 
However, it is possible to tune the area very accurately by adjusting the layout. The similar areas enable optimal point biasing of many qubits
without strongly coupled independent bias lines. 
The experiments reported in this paper are a step towards the tunable coupling of flux qubits.  
The parameters of sample C in particular are very close to those required for the parametric coupling using a third qubit\cite{mypaper}. 
However, for good enough coherence the coupling to the readout 
will have to be modified to be more symmetric than in the present design.

We would like to thank M. Grajcar for useful discussions.

\end{document}